%% file: strong1.tex
\documentstyle[graphics]{europhys}
\input euromacr

\begin{document}
\euro{00}{0}{000-000}{2000}
\Date{00 00 2000}
\shorttitle{A.G. MOREIRA ET AL STRONG-COUPLING THEORY}

\title{ Strong-Coupling Theory for Counter-Ion Distributions}

\author{
Andr\'e G. Moreira and Roland R. Netz}

\institute{
  Max-Planck-Institut f\"ur Kolloid- 
          und Grenzfl\"achenforschung - \\
  Am M\"uhlenberg, 14424 Potsdam, Germany}

\rec {0 00  1999}{}

\pacs{
  \Pacs{82}{70.-y}{Disperse systems}
  \Pacs{61}{20.Qg}{Structure of associated liquids}
  \Pacs{82}{45.+z}{Electrochemistry}
}

\maketitle

\begin{abstract}
The Poisson-Boltzmann approach gives asymptotically exact 
counter-ion density profiles around charged objects
in the weak-coupling limit
of low valency and high temperature. In this paper
we derive, using field-theoretic methods,
 a theory which becomes exact in the opposite
limit of strong coupling. Formally, it corresponds to a 
standard virial expansion. Long-range divergences,
which render the virial expansion intractable for 
homogeneous bulk systems, are shown to be renormalizable 
for the case of inhomogeneous distribution functions
by a systematic expansion in inverse powers of the coupling parameter.
For a planar charged wall, our analytical
results compare quantitatively 
with extensive Monte-Carlo simulations.
\end{abstract}

Recent years have witnessed a revival of the interest in classical
charged systems\cite{Kjellander1,Shklovskii}. 
Specific attention has been paid to the failure of
the Poisson-Boltzmann (PB) 
approach\cite{Kjellander,Attard2,Podgornik,Deserno,Netz2},
which is known to give reliable
results only in the limit of low-valency ions or high temperatures. 
Corrections to PB have been attributed to 
correlations between ions, or, more precisely,
correlated ion-density fluctuations, and, if present,
additional non-electrostatic interactions between ions. 
These corrections are particularly important for the interaction
between macroscopic similarly charged objects, where they can lead to 
attractions\cite{Kjellander,Attard2,Podgornik,Stevens,Lau}.
In as much as the PB approach is accurate
for weakly charged systems, no systematic theory is
available for the distribution of counter-ions around charged
objects in the opposite limit of high-valency ions; moreover, 
it was not clear whether such a limit exists and whether it is
physically meaningful. In this paper we show, using 
field-theoretic methods, that while PB corresponds to the asymptotically
exact theory in the {\em weak-coupling limit},
(corresponding to low-valency ions or high temperatures), our novel
{\em strong-coupling theory} becomes asymptotically exact in the
opposite limit of high-valency ions or low temperatures and constitutes
a physically sound  limit.
For the case of a planar charged wall, we give explicit results
for the asymptotic density profile in the strong-coupling limit. 
We also have performed
extensive Monte-Carlo (MC) simulations of this system. The resulting
density profiles agree for weak and for strong coupling with predictions 
from PB theory and our strong-coupling theory, respectively.
The strong-coupling limit is experimentally easily reached at room
temperatures with highly charged walls and/or multivalent counter ions
and thus relevant from the application point of view.

To proceed, consider the Hamiltonian of a system of $N$ ions
of valency $q$ at an impenetrable and oppositely charged wall 
of number density of surface charges $\sigma_s$,
\begin{equation}
\frac{\cal H}{k_BT} = \sum_{j<k} \frac{\ell_B q^2 }{|{\bf r}_j - {\bf r}_k|}
+ 2\pi q \ell_B \sigma_s  \sum_{j=1}^N z_j
\end{equation}
where $\ell_B = e^2 /4 \pi \varepsilon k_BT$ is the Bjerrum length
which measures the distance at which two unit charges interact
with thermal energy; in water $\ell_B \approx 0.7 nm$.
The dielectric constant is assumed to be homogeneous
throughout the system.
The Gouy-Chapman length, $\mu = 1/2\pi q \sigma_s \ell_B$, 
measures the distance from the wall at which the potential energy 
of an isolated ion reaches thermal energy. Rescaling all
lengths by $\mu$ according to ${\bf r} = \mu \tilde{\bf r}$,
the Hamiltonian reads
\begin{equation} \label{Ham}
\frac{\cal H}{k_BT} = \sum_{j<k}^N \frac{\Xi}
{|\tilde{\bf r}_j - \tilde{\bf r}_k|}
+ \sum_{j=1}^N \tilde{z}_j
\end{equation}
and thus only depends on the coupling parameter 
$\Xi = 2 \pi q^3 \ell_B^2 \sigma_s$.
Using the fact that the typical distance of an ion from the wall is
in reduced units $\tilde{z} \sim 1$ (which holds both in the weak
and strong-coupling limits, as we will show below) the confinement energy,
the second term in Eq.(\ref{Ham}), is of order unity per particle. 
The typical distance between ions scales as 
$\tilde{r} \sim \Xi^{1/2}$ (for $\Xi >1$) or
like $\tilde{r} \sim \Xi^{1/3}$ (for $\Xi<1$ and assuming a liquid-like 
structure) and thus the repulsive energy
between two ions at this distance, as determined by
the first term in Eq.(\ref{Ham}), is of order 
$\Xi^{1/2}$ for $\Xi>1$  or $\Xi^{2/3}$ for $\Xi<1$.
It thus follows on rather general grounds that for
a large coupling parameter, $\Xi > 1$,
the ionic structure is dominated by 
mutual repulsions, suggesting crystallization, while for 
weak coupling $\Xi < 1$ these repulsions should be rather unimportant.
In Fig.1 we show ion-distribution snapshots from Monte-Carlo simulations
for $\Xi=0.1, 10, 10^4$.
For small $\Xi$, Fig.1a, repulsive ion-ion interactions are indeed not
playing a dominant role, the ion distribution is rather disordered 
and mean-field theory should work, since each ion moves
in a diffuse cloud of neighboring ions. For large $\Xi$, on the other hand, 
ion-ion repulsions are strong and ion-ion distances are large 
compared to the distance from  the wall. 
In Fig.1c the ions form a flat layer on the charged wall. A 
two-dimensional one-component plasma is known to crystallize for 
plasma parameter $\Gamma \approx 125$\cite{Baus}. From the
definition of the two-dimensional plasma parameter\cite{Baus}
we obtain the relation $\Xi = 2 \Gamma^2 $. This leads to 
a crystallization threshold (in units of our  coupling parameter) of 
$\Xi \approx 31000$ meaning that all three systems in Fig.1 
are below the crystallization threshold.
However, mean-field theory is expected to break 
down, at least for the system shown in Fig.1c,
because each ion moves, though confined by its immediate neighbors
in the lateral directions, almost independently from the other ions
along the vertical direction (which constitutes the soft mode). 
Let us now see how these
notions can be made concrete in a field-theoretic formulation.
\begin{figure}[t]
\vspace*{-3mm}
\begin{center}
\scalebox{1.00}{\includegraphics{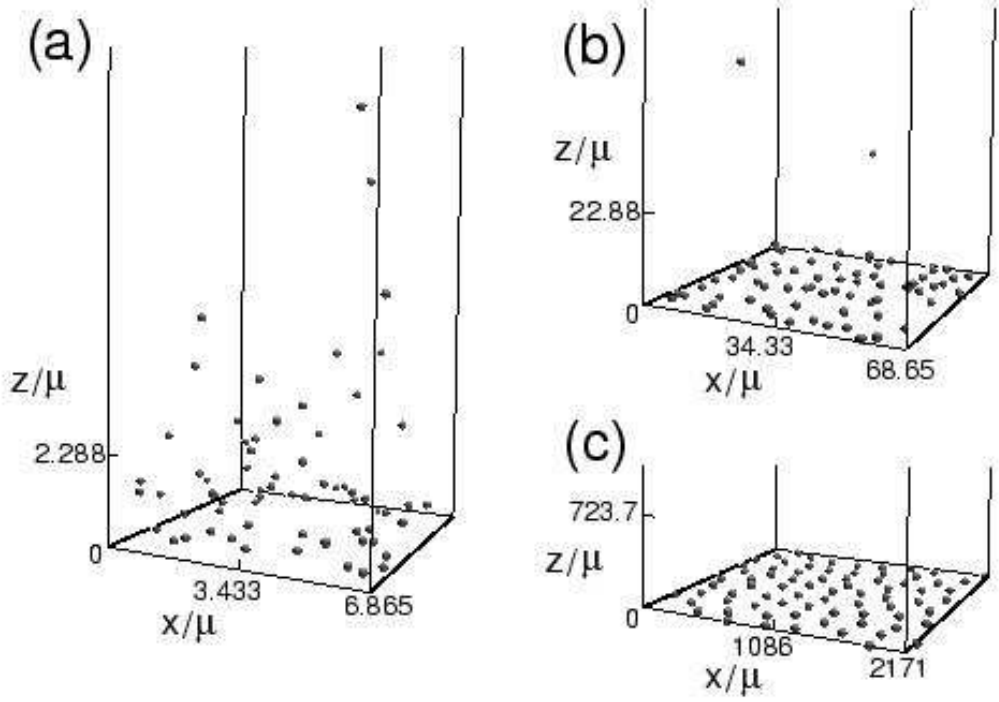}}
%\scalebox{1.00}{\includegraphics{HD:Articles:Figstretch:fig1b.ps}}
\end{center}
\vspace*{-3mm}
\caption{ Snapshots of distributions of 75 ions 
for different values of the coupling constant $\Xi$
from Monte-Carlo simulations.
a) Weak-coupling limit where
Poisson-Boltzmann theory is accurate, $\Xi=0.1$, 
b) intermediate coupling regime, $\Xi = 10$,
and c) strong-coupling limit, $\Xi = 10000$. 
Note that in a) 8 particles which are located far away from the charged
surface are not shown.}
\vspace*{-3mm}
\end{figure}

The partition function of a system of $N$ counter-ions 
interacting via Coulomb interactions $v({\bf r}) = 1/r$
among themselves and with a distribution $\sigma({\bf r})$ 
is given by
\begin{equation}
{\cal Z  }_N = \frac{1}{N!} \prod_{j=1}^N \int {\rm d} {\bf r}_j
\exp \left\{ -q^2 \ell_B \sum_{j<k} v({\bf r}_j - {\bf r}_k) 
+q \ell_B \int {\rm d}{\bf r} \sigma({\bf r}) \sum_j v({\bf r}-{\bf r}_j) 
+ \sum_j h({\bf r}_j) \right\} \nonumber
\end{equation}
where the field $h$ has been added to calculate density distributions
later on. We implicitly assume the configurational integral to
span the upper half space ($z>0$) only.
At this point we employ a Hubbard-Stratonovitch transformation,
similar to previous implementations of a field theory for charged 
system\cite{Netz2,Netz1}, which leads to
\begin{equation}
{\cal Z  }_N = \frac{1}{N!} \int 
\frac{{\cal D}\phi }{{\cal Z}_v} 
\exp \left\{ -\frac{1}{2\ell_B } \int {\rm d}{\bf r}{\rm d}{\bf r}'
\phi({\bf r}) v^{-1}({\bf r}-{\bf r}') \phi({\bf r}')
+ \imath \int {\rm d}{\bf r} \sigma({\bf r}) \phi({\bf r}) \right\}
\left[ \int {\rm d} {\bf r} {\rm e}^{h({\bf r}) -\imath q \phi({\bf r})}
\right]^N \nonumber
\end{equation}
where we introduced the notation ${\cal Z}_v = \sqrt{\det v}$.
For simplicity, we perform a Legendre transformation  to the
grand-canonical ensemble,
${\cal Q} = \sum_N \lambda^N {\cal Z}_N$, and thereby introduce the fugacity
$\lambda$. The inverse Coulomb operator follows from
Poisson's law as
$v^{-1}({\bf r}) = -\nabla^2 \delta({\bf r})/4 \pi$. The 
grand-canonical partition function can therefore be written as
\begin{equation}
{\cal Q  } =  \int \frac{{\cal D}\phi }{{\cal Z}_v} 
\exp \left\{ - \int {\rm d}{\bf r} \left[
\frac{ [\nabla \phi({\bf r})]^2}{8 \pi \ell_B}
- \imath \sigma({\bf r}) \phi({\bf r})
-\lambda  \theta(z) 
{\rm e}^{h({\bf r}) -\imath q \phi({\bf r})} \right] \right\}
\end{equation}
where $\theta(z) =1$ for $z>0$ and zero otherwise.
Next we rescale the action, similarly to our rescaling analysis of the
Hamiltonian in the beginning. All lengths are rescaled by the 
Gouy-Chapman length, ${\bf r} = \mu \tilde{\bf r}$, 
the fluctuating field is rescaled by the valency, $\phi = \tilde{\phi}/q$.
We also use the explicit form $\sigma({\bf r}) = \sigma_s \delta (z)$
for the fixed charge distribution. The result is
\begin{equation} \label{Q}
{\cal Q  } =  \int \frac{{\cal D} \tilde{\phi} }{{\cal Z}_v} 
\exp \left\{ - \frac{1}{8\pi \Xi} \int {\rm d}\tilde{\bf r} \left[
[\nabla \tilde{\phi}(\tilde{\bf r})]^2
-4 \imath \delta(\tilde{z}) \tilde{\phi}(\tilde{\bf r})
-4\Lambda  \theta(\tilde{z}) {\rm e}^{h(\tilde{\bf r}) -
\imath \tilde{\phi}(\tilde{\bf r})} \right] \right\}
\end{equation}
where the rescaled fugacity $\Lambda$ is defined by
\begin{equation}
\Lambda = 2 \pi \lambda \mu^3 \Xi = 
\frac{\lambda}{2 \pi \sigma_s^2 \ell_B}.
\end{equation}
The expectation value of the
counter-ion density, $\langle \rho(\tilde{\bf r}) \rangle$, follows 
 by taking a functional derivative with respect
to the generating field $h$, $\langle \rho(\tilde{\bf r}) \rangle =
\delta \ln {\cal Q} / \delta h(\tilde{\bf r})  \mu^3$,
giving rise to
\begin{equation} \label{dens}
\frac{\langle \rho(\tilde{\bf r}) \rangle }{2 \pi \ell_B \sigma_s^2}=
 \Lambda \langle {\rm e}^{-\imath
\tilde{\phi}(\tilde{z})} \rangle.
\end{equation}
The normalization condition for the counter-ion distribution,
$\mu \int {\rm d} \tilde{z} \rho(\tilde{z}) = \sigma_s/ q$,
which follows directly from the definition of the grand-canonical 
partition function, leads to
\begin{equation} \label{norm}
\Lambda \int_0^\infty {\rm d} \tilde{z} \langle {\rm e}^{-\imath
\tilde{\phi}(\tilde{z})} \rangle = 1.
\end{equation}
This is an important equation since it shows that the
expectation values of the
fugacity term in Eq.(\ref{Q}) is bounded and of the order
of unity per unit area.
Let us first repeat the saddle-point analysis, which, because
of the structure of the action in Eq.(\ref{Q}), should be valid for
$\Xi \ll 1$. The saddle-point equation reads
\begin{equation} \label{DG}
\frac{ {\rm d}^2 \tilde{\phi}(\tilde{z})}{{\rm d} \tilde{z}^2} =
2 \imath \Lambda {\rm e}^{ - \imath \tilde{\phi}(\tilde{z})}
\end{equation}
with the boundary condition ${\rm d} \tilde{\phi}(\tilde{z}) /
{\rm d} \tilde{z} = -2 \imath$ at $\tilde{z}=0$. 
The solution of this differential
equation is
\begin{equation} \label{PB}
\imath \tilde{\phi}(\tilde{z}) = 2 \ln \left(1 + \Lambda^{1/2} 
\tilde{z} \right)
\end{equation}
while the boundary condition leads to $\Lambda =1$,
which shows that the saddle-point approximation is indeed valid
in the limit $\Xi \ll 1$.
Combining Eqs.(\ref{dens}) and (\ref{PB}),
the density distribution of counter ions is given by the well-known result
\begin{equation} \label{PBdens}
\frac{ \langle \rho(\tilde{\bf r}) \rangle}{2 \pi \ell_B \sigma_s^2} = 
\frac{1}{(1 + \tilde{z})^2}.
\end{equation}
Let us now consider the opposite limit, when the 
coupling constant $\Xi$ is large. In this case, the 
saddle-point approximation breaks down, since the prefactor in front
of the action in Eq.(\ref{Q}) becomes small. 
This has been seen in a systematic 
loop-wise expansion around the saddle point, where the
corrections to the saddle-point solution are
proportional to powers of $\Xi$ and thus become large\cite{Netz2}. 
However, from the field-theoretic partition function Eq.(\ref{Q}), 
it is self-evident what has to be done in this limit. Since the fugacity
term is bounded, as evidenced by Eq.(\ref{norm}), one can
expand the partition function (and also all expectation values)
in powers of $\Lambda/\Xi$. 
Upon Legendre transformation to the canonical ensemble, this gives 
the standard virial expansion.
The normalization condition Eq.(\ref{norm}) 
can be solved by an expansion of the fugacity as
$\Lambda = \Lambda_0 + \Lambda_1 /\Xi + \ldots$, which 
leads to an expansion of the density profile with
the small parameter $1/\Xi$.
While the standard virial expansion fails for homogeneous bulk systems
because of infra-red divergences, these divergences are renormalized
for the present case of inhomogeneous distribution functions via 
the normalization condition Eq.(\ref{norm}) as we will now demonstrate.  
To leading order in this expansion, the rescaled density is 
\begin{equation}
\frac{\langle \rho(\tilde{\bf r}) \rangle }{2 \pi \ell_B \sigma_s^2}=
 \Lambda \exp\left\{-\frac{\Xi}{2} v(0) 
 +\frac{1}{2 \pi} \int {\rm d}{\bf r}
 v({\bf r} - \tilde{\bf r})  \delta(z) \right\}.
\end{equation}
From the normalization condition Eq.(\ref{norm}) we obtain
\begin{equation}
 \Lambda_0=  \exp\left\{\frac{\Xi}{2} v(0) 
 -\frac{1}{2 \pi} \int {\rm d}{\bf r}
 v({\bf r})  \delta(z) \right\}
\end{equation}
and thus the density distribution is to leading order given by
\begin{equation} \label{strong}
\frac{ \langle \rho(\tilde{\bf r}) \rangle}{2 \pi \ell_B \sigma_s^2} = 
{\rm e}^{\displaystyle - \tilde{z}}.
\end{equation}
\begin{figure}[t]
\vspace*{-3mm}
\begin{center}
\scalebox{0.60}{\includegraphics{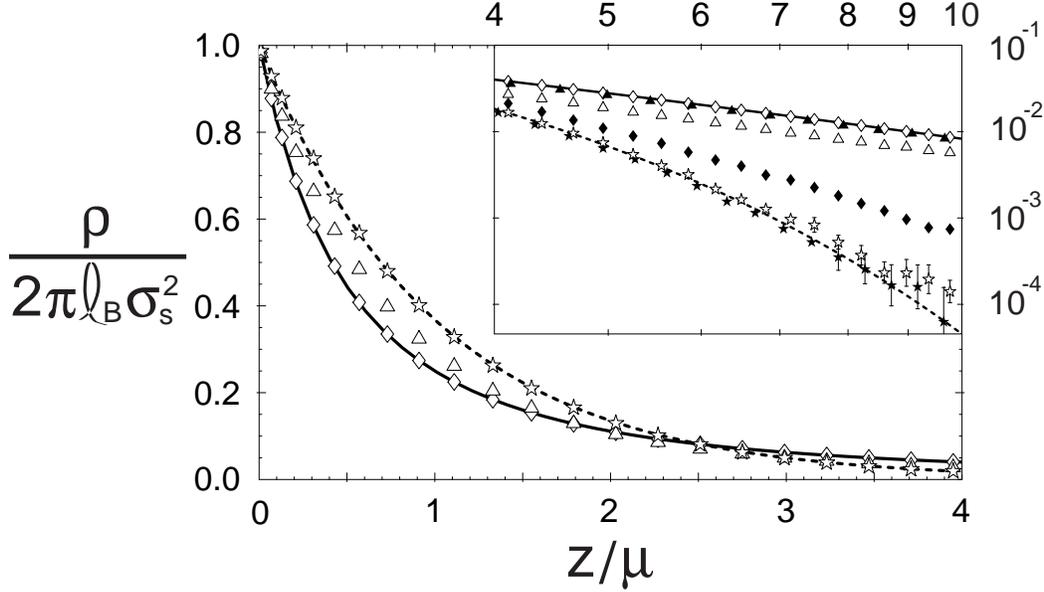}}
%\scalebox{1.00}{\includegraphics{HD:Articles:Figstretch:fig1b.ps}}
\end{center}
\vspace*{-3mm}
\caption{ Rescaled counter-ion density distribution 
$\rho/2\pi \ell_B \sigma_s^2 $
as a function of the rescaled distance from the wall $z/\mu$. 
The inset shows Monte Carlo data for coupling constants 
$\Xi = 10^5, 10^4, 100, 10, 1, 0.1$ 
(from bottom to top) in a double-logarithmic plot, the main figure
shows data for $\Xi = 0.1$ (open diamonds), $\Xi = 10$ (open
triangles) and $\Xi = 10^4$ (open stars). The solid and broken lines 
denote the Poisson-Boltzmann and strong-coupling predictions,
Eqs.(\ref{PBdens}) and (\ref{strong}), respectively.
All data were obtained with 75 particles and $10^6$ Monte Carlo
steps (MCS), except the 
data for $\Xi=0.1$ where 600 particles were simulated. Error bars
are smaller than the symbol size if not shown.}
\vspace*{-3mm}
\end{figure}
The density profile in the strong coupling limit $\Xi \rightarrow \infty$
is thus given by a simple exponential, while the profile
in the weak-coupling (Poisson-Boltzmann) limit $\Xi \rightarrow 0$,
Eq.(\ref{PBdens}), is given by a power law. Both
distributions give the same density at the wall, which is a trivial
consequence of the contact-value theorem since the net force on
the wall is zero\cite{Netz2}.
Just to avoid confusion at this stage, we stress that the
exponential density profile Eq.(\ref{strong})
has nothing to do with the Debye-H\"uckel approximation.
It is true that an exponential density profile 
(though with a different density contact value and thus
violating the contact-value theorem)
also follows from linearizing the differential equation Eq.(\ref{DG}).
However, the linearized solution can never be a more faithful 
representation of the true density profile than the full non-linear 
solution, and secondly, nothing in the saddle-point equation indicates
that linearization should be valid for large values of $\Xi$, 
because Eq.(\ref{DG})  does not depend on $\Xi$. 
It is the saddle-point approach itself (the framework
within which the Debye-H\"uckel approximation can be formulated) 
which becomes invalid when $\Xi $ becomes large.

In Fig.2 we show counter-ion density profiles obtained using
Monte Carlo simulations for various values of the coupling parameter
$\Xi$. As can be seen, the PB density profile Eq.(\ref{PBdens}) 
is only realized for $\Xi <1$, while  the strong-coupling profile
Eq.(\ref{strong}) is indeed the asymptotic solution and
agrees with simulation results for $\Xi > 10^4$. 
Experimentally, a coupling parameter $\Xi= 100$, which is already
quite close to the strong-coupling limit (see Fig.2 inset),
is reached with divalent ions for a surface charged density
$\sigma_s \approx 3.6 nm^{-2}$, which is feasible with compressed
charged monolayers, and with trivalent counter ions for 
$\sigma_s \approx 1 nm^{-2}$, which is a typical value.
The strong-coupling limit is therefore experimentally accessible.
\begin{figure}[t]
\vspace*{-3mm}
\begin{center}
\scalebox{0.50}{\includegraphics{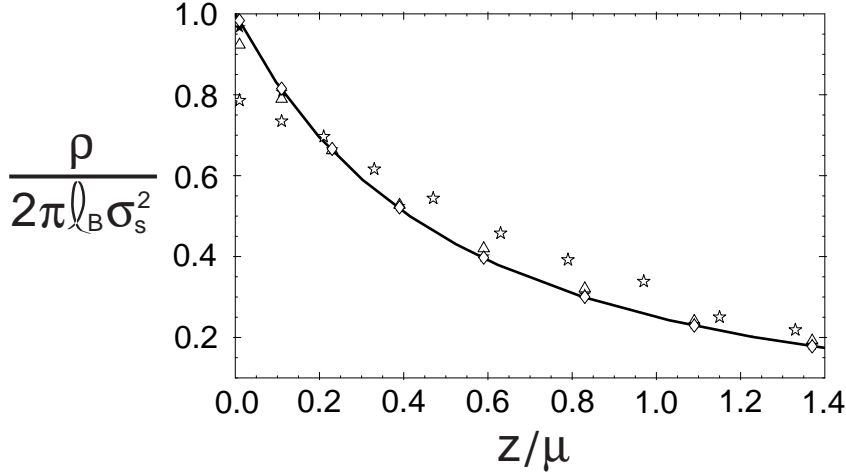}}
%\scalebox{1.00}{\includegraphics{HD:Articles:Figstretch:fig1b.ps}}
\end{center}
\vspace*{-3mm}
\caption{ Results for the rescaled counter-ion 
density distribution $\rho/2\pi \ell_B \sigma_s^2 $
as a function of the rescaled distance from the wall $z/\mu$
for   $\Xi=0.1$ and
5 particles and $10^8$ MCS (open stars), 
15 particles and $10^7$ MCS (open triangles), 
35 particles and $10^7$ MCS (filled stars), and 
600 particles and $10^6$ MCS (open diamonds). 
Finite size are negligible except for 
the simulations using 5 particles and (to a lesser degree)
15 particles.}
\vspace*{-3mm}
\end{figure}
In our simulations we employ periodic boundary conditions in the
lateral directions, while the ions are unconfined vertically.
Forces and energies are
calculated following the approach by Lekner\cite{Lekner}
and Sperb\cite{Sperb} where the sum over the periodic images
is transformed into a rapidly converging expansion in terms
of Bessel functions\cite{Felix}. Since the number of particles
in the simulation, $N$, is related to the linear box size
(in units of the Gouy-Chapman length $\mu$)  by
$\tilde{L} = L/\mu = \sqrt{2 \pi N \Xi}$, effects due to the finite
number of particles and effects due to the finite lateral box size
are connected. In Fig.3 we show MC density profiles for a fixed
coupling constant $\Xi=0.1$ and for various number of particles. 
For all but the smallest systems with $5$ and $15$ particles
finite-size effects are negligible. 

In summary, we have derived within a field-theoretic framework
the strong-coupling theory for counter-ion distributions,
valid in the limit of large valencies and/or low temperatures.
It corresponds to a standard virial expansion of the field-theoretic action, 
which gives meaningful results because long-range divergences,
which spoil the free-energy virial expansion for bulk systems
and lead to non-analytic terms\cite{Netz3}, are subtracted by a 
fugacity renormalization. 
This theory  thereby complements the 
Poisson-Boltzmann approach, which works for the oppposite limit
of weak coupling.
We note that an exponential density profile
(albeit with a different prefactor) has been derived in 
the strong-coupling limit by Shklovskii using 
a heuristic model where ions bound to the wall are in chemical
equilibrium with free ions\cite{Shklovskii}.

Next-leading order results for the density profile within 
our strong-coupling theory, and comparison of these corrections
with Monte-Carlo simulations,  will be presented in a future 
publication. Also, generalization to other geometries, 
including the pressure between two charged walls, the
 presence of salt, dielectric boundaries, and hard-core interactions
are straightforward within our theory and will be published soon. 
For the case of a cylinder with a radius $R$ and linear charge density
$\tau$ we obtain that the strong-coupling theory is valid for 
$\Xi_c \gg 1 $ with the cylindrical coupling parameter defined as
$\Xi_c = q^3 \ell_B^2 \tau/R$. 
The Manning condensation threshold is reflected by the fact that the 
density distribution is not normalizable for $\tau \ell_B <1$. 
For $\tau \ell_B >1$  only part of the counter ions are
condensed and the normalization condition Eq.(\ref{norm})
is modified.
The result for the density profile
in the limit $\Xi_c \gg 1 $  is 
\begin{equation}
\rho(r) = \frac{(\tau \ell_B q -1)^2}{\pi \ell_B q^2 R^2}
\left( \frac{r}{R}\right)^{-2 \tau \ell_B q}
\end{equation}
and thus exhibits purely algebraic decay, in contrast to the 
PB prediction\cite{Netz4} but in agreement with Shklovskii's
results\cite{Shklovskii}.
Finally, we note that the strong-coupling limit not 
necessarily entails unrealistically small Gouy-Chapman
lengths. Writing the Gouy Chapman length as
$\mu = \ell_B q^2 / \Xi$, it is clear that
$\mu$ can be fixed at a moderate value for highly
valent ions even when $\Xi \gg 1$. In this case,
however, one would have to worry about the effect of
discrete surface charges.

\stars

We thank Felix Csajka, Henri Orland, 
Christian Seidel, and Julian Shillcock
for useful discussions.
AGM  acknowledges financial support from FCT
through the grant Praxis XXI/BD/13347/97.

\bibliographystyle{}

\end{document}

%% file: euromacr.tex
\def\stars{\bigskip\centerline{***}\medskip}

\newif\ifboo \boofalse